\documentclass{article} 
\usepackage[preprint,nonatbib]{neurips_2022}

\usepackage{amsmath,amsfonts,bm}

\def\eqref#1{equation~\ref{#1}}

\def\1{\bm{1}}

\DeclareMathAlphabet{\mathsfit}{\encodingdefault}{\sfdefault}{m}{sl}
\SetMathAlphabet{\mathsfit}{bold}{\encodingdefault}{\sfdefault}{bx}{n}

\usepackage{hyperref}
\usepackage{url}
\usepackage{amsmath} 
\usepackage{booktabs} 
\usepackage{multirow}
\newcommand{\angstrom}{\text{\normalfont\AA}}
\usepackage{enumitem}
\usepackage{graphicx}
\usepackage{subfigure}
\usepackage{xcolor}
\usepackage{tablefootnote} 

\setlist{topsep=0pt, leftmargin=15pt}

\title{
Do Deep Learning Models Really Outperform Traditional Approaches in Molecular Docking?
}

\author{Yuejiang Yu, Shuqi Lu, Zhifeng Gao, Hang Zheng \& Guolin Ke \\
DP Technology\\
\texttt{\{yuyj, lusq, gaozf, zhengh, kegl\}@dp.tech} \\
}

\begin{document}

\maketitle

\begin{abstract}
Molecular docking, given a ligand molecule and a ligand binding site (called ``pocket'') on a protein, predicting the binding mode of the protein-ligand complex, is a widely used technique in drug design.
Many deep learning models have been developed for molecular docking, while most existing deep learning models perform docking on the whole protein, rather than on a given pocket as the traditional molecular docking approaches, which does not match common needs. 
What's more, they claim to perform better than traditional molecular docking, but the approach of comparison is not fair, since traditional methods are not designed for docking on the whole protein without a given pocket.
In this paper, we design a series of experiments to examine the actual performance of these deep learning models and traditional methods. 
For a fair comparison, we decompose the docking on the whole protein into two steps, pocket searching and docking on a given pocket, and build pipelines to evaluate traditional methods and deep learning methods respectively. 
We find that deep learning models are actually good at pocket searching, but traditional methods are better than deep learning models at docking on given pockets. 
Overall, our work explicitly reveals some potential problems in current deep learning models for molecular docking and provides several suggestions for future works. 

\end{abstract}





\section{Introduction}
Molecular docking, given a ligand molecule and a ligand binding site (called ''pocket'') on a protein, predicting the binding mode of the protein-ligand complex, is a widely used technique in drug design, and several molecular docking methods ~\cite{trott2010autodock, alhossary2015fast_qvina2, hassan2017protein_qvinaw, quiroga2016vinardo} are proposed in the last decades. 

Recently, deep learning has been widely used in drug design applications, such as molecular property prediction~\cite{zhou2022uni,rong2020self,wang2022molecular,fang2022geometry} and protein structure predition~\cite{jumper2021highly}. Several recent works~\cite{stark2022equibind,lu2022tankbind,corso2022diffdock} also try to apply deep learning to molecular docking. 
However, most of them did not follow the setting of traditional molecular docking. In particular, instead of docking on a given pocket, these deep learning works directly perform docking on the whole protein (called ``blind docking'' \footnote{We use ``molecular docking'' to refer to the docking on a given pocket in this paper.}), which does not match common needs in drug design. 
Besides, although they claimed they are better than traditional molecular docking approaches, the experiment and the evaluation in their papers are questionable. Specifically, we have the following questions. 
\begin{itemize}
    \item \emph{Are the comparisons with traditional approaches fair?} The traditional molecular docking approaches are not designed for blind docking. But the early works \cite{stark2022equibind,lu2022tankbind,corso2022diffdock} directly apply traditional molecular docking approaches on the whole proteins, rather than the pockets, in the experiments. It is obvious that such a comparison is not fair for traditional molecular docking approaches.
    \item \emph{What are these deep learning models really good at? Pocket searching or molecular docking?} 
    In traditional methods, blind docking is generally divided into two steps: pocket searching and docking. Although these works combine pocket searching and molecular docking together, they simply evaluate the performance of the final complex conformation. We do not know whether the gain is brought by better pocket searching or better molecular docking. Thus, we actually do not know whether these models can outperform the traditional approaches in molecular docking or not, given the same protein pockets.
\end{itemize}

To answer the above questions, we design a series of experiments in Sec.~\ref{sec:experiments}, and have the following findings.
\begin{itemize}
    \item \emph{Traditional molecular docking approaches are still better than deep learning models, when given the same pockets.} Besides, the two-stage traditional approaches (traditional pocket searching + molecular docking) can outperform most of the deep learning models.
    \item \emph{Deep learning models are actually good at pocket searching.} Although their performance on molecular docking is worse, deep blind docking models actually perform well in pocking searching. Besides, we find there is still a large room to reach the performance upper bound of pocket searching.
    \item  { \emph{Deep Learning models are full of potential.} DiffDock~\cite{corso2022diffdock} is currently the best method in pocket searching, and its performance in molecular docking is very close to traditional approaches. It is no doubt that deep learning models will outperform traditional approaches in the future. 
    }
\end{itemize}

\section{Related Work}

\label{Related work}

\paragraph{Pocket Searching Approaches} Several pocket searching approaches are proposed to find possible new biological targets on proteins. Fpocket~\cite{le2009fpocket} includes pocket finding, ranking and describing based on Voronoi tessellation and alpha spheres. P2Rank~\cite{krivak2018p2rank} is a convolutional neural network based tool. PointSite~\cite{pointsite2022} further leverages the 3D spatial information, based on 3D U-Net model~\cite{cciccek20163d}. 

\paragraph{Traditional Molecular Docking Approaches}  
AutoDock and its variants ~\cite{trott2010autodock,ravindranath2015autodockfr,eberhardt2021autodock,santos2021accelerating} are widely used in molecular docking. Thanks to Vina and AutoDock4 scoring function and fast search methods, binding poses and affinities of ligands are predicted efficiently. Based on AutoDock, Vinardo~\cite{quiroga2016vinardo}, QVina2~\cite{alhossary2015fast_qvina2}, QVina-W~\cite{hassan2017protein_qvinaw}, and Smina~\cite{koes2013lessons_smina} are developed to further improve searching and scoring power of docking. To better perform virtual screening over ultra-large ligand datasets, GPU-accelerated docking engines including Uni-Dock~\cite{yu2022uni} and AutoDock-GPU~\cite{santos2021accelerating} broaden the throughout of docking dramatically.

\paragraph{Deep Learning Models for Docking} Recent advances in deep learning-based molecular docking focus on blind docking, which have zero knowledge of the protein’s pocket and directly predict the molecular positions and conformations. EquiBind ~\cite{stark2022equibind} proposed a SE(3)-equivariant geometric deep learning model to predict the molecular positions and conformations by directly predicting the 3D atom coordinates of molecules. Similarly, several later works~\cite{zhang2022e3bind, lu2022tankbind, corso2022diffdock} also focus on blind docking.
TANKBind \cite{lu2022tankbind} proposes a two-stage deep docking framework, which segments the whole protein into functional blocks and predicts their interactions with the ligand using a trigonometry-aware architecture, and then the binding structure is prioritized according to the predicted interactions.
DiffDock~\cite{corso2022diffdock} randomly samples a molecular conformation and predicts the molecular atom coordinates  with a denoising diffusion probability model starting from the random molecular conformation.








\section{Experiments}
\label{sec:experiments}


We design a series of experiments, to have a fair comparison between traditional molecular docking approaches and deep learning based models. In particular, since traditional approaches cannot search pockets, we introduce additional pocket searching tools, or directly use the pockets found by the deep learning models, for them. 
The details of the experimental designs are in the following subsection.

\subsection{Experimental Settings}


\paragraph{Data} We use the same benchmark dataset as Equibind~\cite{stark2022equibind} and DiffDock~\cite{corso2022diffdock}. The test set is collected from PDBBind~\cite{liu2017forging}. The conformations of molecular ligands are initialized by RDKit, and the protein structures are taken from \hyperlink{https://anonymous.4open.science/r/DiffDock}{https://anonymous.4open.science/r/DiffDock}. 

\paragraph{Deep learning models} The below models are used as baselines.
\begin{itemize}
    \item EquiBind~\cite{stark2022equibind}, a docking tool that predict the conformation without binding site knowledge. We directly use the number reported in DiffDock~\cite{corso2022diffdock} paper.
    \item TANKBind~\cite{lu2022tankbind}, we use the number reported in DiffDock~\cite{corso2022diffdock} paper.
    \item TANKBind*, our reproduced TANKBind's results, based on its officially released model weights and source codes ~\footnote{\url{https://github.com/luwei0917/TankBind}}. We run the model three times with different random seeds, and report the mean and std like in DiffDock ~\cite{corso2022diffdock}.
    \item DiffDock~\cite{corso2022diffdock}, a generative model of binding pose prediction of  under the setting from the corresponding paper ~\cite{corso2022diffdock}, which generates 40 samples with 20 diffusion steps. We directly use the reported number in their paper.
    \item DiffDock*, our reproduced DiffDock's results, based on its officially released model weights and source codes ~\footnote{\url{https://github.com/gcorso/DiffDock}}.  We run the results three times with different random seeds, and report the mean and std the same way as TANKBind*.
\end{itemize}


\paragraph{Traditional approaches} To have a fair comparison with deep learning models in the blind docking setting, we applied the two-stage solution for the traditional approaches: pocket searching, then molecular docking. For the molecular docking, we use Uni-dock~\cite{yu2022uni}, a GPU-accelerated docking tool, for its high efficiency. Then, with different pocket searching methods, we totally set up the following 5 configurations. 
\begin{itemize}
\item 1. Fpocket + Uni-dock, which uses pockets found by Fpocket~\cite{le2009fpocket}. In particular, we directly use the pocket with the best ``fpocket score''. Using the geometric center of the predicted pocket atoms as the center, we create an axis-parallel cube with 30 $\angstrom$ edge size for molecular docking. 
\item 2. P2Rank + Uni-dock, which uses pockets found by P2Rank~\cite{krivak2018p2rank}. In particular, we directly use the rank-1 pocket predicted by P2Rank. Using the geometric center of the predicted pocket atoms as the center, we create an axis-parallel cube with 30 $\angstrom$ edge size for molecular docking. 
\item 3. PointSite + Uni-dock, which uses PointSite~\cite{pointsite2022}, a 3D U-Net model to find the pocket atoms. PointSite only predicts one pocket for one protein, and we directly use it. Using the geometric center of the predicted pocket atoms as the center, we create an axis-parallel cube with 30 $\angstrom$ edge size for molecular docking. 
\item 4. DiffDock* + Uni-dock, which uses pockets found by DiffDock. In particular, based on our reproduced DiffDock*, for each protein, we select the top-1 conformation based on DiffDock's predicted confidence scores. Then, we create a minimal axis-parallel rectangular cuboid that can cover all ligand molecular atoms. Finally, we enlarge the cuboid in three axes by 5 $\angstrom$, and use the enlarged cuboid for molecular docking. Note that here we only use the pockets found from DiffDock in this setting, rather than its predicted molecular conformations.
\item 5. GT pocket + Uni-dock, which uses the ground-truth (GT) pockets directly. Similar to DiffDock's pockets, we create a minimal axis-parallel rectangular cuboid that can cover all ground-truth molecular atoms. Then, we enlarge the cuboid in three axes by 5 $\angstrom$, and use the enlarged cuboid for molecular docking. We set up this experiment to demonstrate the performance upper bound of traditional docking approaches when given the correct pockets.
\end{itemize}

All the above settings are run three times by different random seeds, and we report the means and standard deviations for evaluation metrics. We also provide a repository \footnote{\url{https://github.com/pkuyyj/Blind\_docking}} for reproducing our results. Notably, in the repository, we use open-sourced Autodock Vina as the traditional docking engine, since Uni-dock is not open-sourced yet. Despite this, the repository should be able to reproduce our results, as Uni-dock can be recognized as the GPU-accelerated version of Autodock Vina.

\paragraph{Evaluation metrics} The evaluation also follows DiffDock~\cite{corso2022diffdock}. In particular, we first select the top-$k$ ($k \in \{1, 5\}$) conformations based on docking scores (or confidence scores predicted by deep learning models), for each protein. Then, we compute the heavy atoms' RMSD for them, based on ground-truth conformations. Next, we select the best conformation with the minimal RMSD for each protein. Finally, we report the percentage of best conformation with RMSD $< m\angstrom$ ($m \in \{1, 2\}$) on all proteins, and the median RMSD of the best conformation on all proteins. Notably, the RMSD between two conformations considers the symmetry permutations, the same as DiffDock's evaluation.




\subsection{Results}

\paragraph{Blind docking performance}

From the end-to-end blind docking performance shown in Table~\ref{tab:overall_results}, we have the following findings.
1) For the accurate docking conformation (percentage of RMSD $< 1\angstrom$), traditional approaches (with P2Rank and PointSite) can outperform deep learning approaches. 2) Even for less accurate metrics (percentage of RMSD $< 2\angstrom$), traditional approaches still outperform EquiBind and TankBind. 3) DiffDock's performance on RMSD $< 2\angstrom$ is good, but Uni-Dock with DiffDock's pocket is even better. This indicates its performance is gained from the better pocket searching ability.


\paragraph{Molecular docking performance}
We can compare DiffDock* with ``DiffDock* +Uni-dock'', to examine the performance of molecular docking, when given the same pockets. From the results, it is clear that ``DiffDock* +Uni-dock'' consistently outperforms DiffDock*. This shows when using the same pocket, traditional molecular docking approaches are still better than deep learning models.

\paragraph{Pocket searching performance}
We can compare Uni-dock with different pocket searching methods, to examine the pocket searching performance. 1) It is easy to find that PointSite is the best, and fpocket is the worst. 2) The pockets found by DiffDock* are quite good, outperforming all existing pocket searching tools. We suppose the gain is from the additional molecule inputs used in DiffDock, since pfocket, P2Rank, and PointSite only take the protein as input.
3) When using the ground-truth pockets, the performance of Uni-Dock largely outperforms all other methods. This indicates there is still a large room to reach the performance upper bound of pocket searching.


\begin{table}[t]
\centering
\small
  \caption{Performance of blind docking.}
  \label{tab:overall_results}
    \scalebox{0.85}{
    \addtolength{\tabcolsep}{-2pt}
  \begin{tabular}{l|l|ccc|ccc}
    \toprule
    & \multirow{2}{*}{Method} & \multicolumn{3}{c|}{Top-1 RMSD(\angstrom)} &  \multicolumn{3}{c}{Top-5 RMSD(\angstrom)} \\
     & & $\% < 1\angstrom$ ($\uparrow$) & $\% < 2\angstrom$ ($\uparrow$) & Med. ($\downarrow$) & $\% < 1\angstrom$ ($\uparrow$) & $\% < 2\angstrom$ ($\uparrow$) & Med. ($\downarrow$) \\
    \midrule
     \multirow{2}{*}{Deep Learning} &EquiBind & - & 5.5$\pm$1.2 & 6.2$\pm$0.3 & -& - & - \\
     &TANKBind & &20.4$\pm$2.1 & 4.0$\pm$0.2 && 24.5$\pm$2.1 & 3.4$\pm$0.1 \\
     &TANKBind* & 2.66$\pm$0.26 &18.18$\pm$0.6& 4.2$\pm$0.05 &4.13$\pm$0.0& 20.39$\pm$0.45 & 3.5$\pm$0.04 \\
     &DiffDock & &38.2$\pm$2.5 & 3.30$\pm$0.3 && 44.7$\pm$2.6 & 2.40$\pm$0.2   \\
     &DiffDock*  & 15.41$\pm$0.49\footnotemark[3] &36.62$\pm$0.35 & 3.31$\pm$0.03 & 21.58$\pm$0.38\footnotemark[3] & 44.19$\pm$0.49 & 2.37$\pm$0.06 \\
     \midrule
    \multirow{2}{*}{Traditional}&Fpocket + Uni-dock  & 13.33$\pm$0.4 &18.7$\pm$0.13 & 13.2$\pm$0.26 & 19.16$\pm$0.39 & 27.32$\pm$0.69 & 8.3$\pm$0.25 \\
    &P2Rank + Uni-dock &19.31$\pm$1.07 & 28.6$\pm$1.17 & 6.4$\pm$0.22 & 27.76$\pm$1.03 & 39.18$\pm$1.03 & 3.76$\pm$0.06  \\
    &PointSite + Uni-dock &21.36$\pm$1.65& 32.12$\pm$0.93 & 5.54$\pm$0.46 &31.38$\pm$0.86& 46.06$\pm$0.69 & 2.52$\pm$0.18 \\
    \midrule
    Better Pocket &DiffDock* + Uni-dock  & 25.49$\pm$0.6 &  38.93$\pm$0.23 & 4.14$\pm$0.07 & 36.97$\pm$1.05 & 51.07$\pm$1.06 & 1.93$\pm$0.12 \\
    + Traditional & GT pocket + Uni-dock & 32.77$\pm$0.38 & 51.11$\pm$0.6 & 1.89$\pm$0.04 & 47.5$\pm$0.23 & 67.59$\pm$0.94 & 1.11$\pm$0.02 \\
    \bottomrule
  \end{tabular}
}
\end{table}
\footnotetext[3]{Here, we use the confidence model provided in https://github.com/gcorso/DiffDock to select molecular conformations with RMSD $< 1\AA$. Although the confidence model here is trained based on molecular conformations with RMSD $< 2\AA$, it can also reflect the model performance to some extent. We are working on training a DiffDock with confidence model RMSD $< 1\AA$.}


\subsection{Discussions}
According to the above experimental results, we can conclude that current deep learning is actually good at pocket searching, not molecular docking. We have the following suggestions for future deep learning models for molecular docking:
\begin{itemize}
    \item Focus on molecular docking with given pockets, rather than blind docking. In real-world applications, pockets are usually known and fixed in a drug design project.
    \item Pocket searching itself is an important problem, and still has a large room for improvement. Besides, if the additional ligand molecules can be used as model inputs (like in DiffDock), the performance may be further improved.
    \item For end-to-end blind docking models, the comparison in experiments should be fair. In particular, you should first use pocket searching approaches (or directly use the same pockets as your models), then applied the traditional molecular docking methods in the pockets, rather than in the whole proteins. 
    \item { The deep learning models are full of potential, especially DiffDock. From the results, we can find that DiffDock is the best model in pocket searching, and achieve almost comparable performance with traditional approaches. We believe the deep learning models could be further improved in the near future.}
\end{itemize}

    

\section{Conclusion}
Several deep learning models have been proposed for molecular docking. However, these models focus on blind docking, different from the docking in a given pocket in traditional approaches. Besides, in their experiments, the comparison with traditional approaches is usually not fair, i.e., they use the traditional approaches in the whole protein, rather than in a given pocket. To examine the actual performance of deep learning models, we design a series of experiments to compare traditional molecular docking approaches with deep learning based models. Our experimental results indicate that, traditional molecular docking approaches still outperform deep learning models, when using the same pockets. 
Based on our findings, we suggest the community and future works on molecular docking can correctly evaluate the traditional approaches. Besides, since blind docking actually does not align with { common }real-world applications, it is better to solve pocket searching and molecular docking (on given pockets) separately.




\subsection*{Acknowledgments}
We thank Gengmo Zhou, Junhan Chang and many colleagues in DP Technology for their great help in this project.
Most experiments are carried out on the cloud platform Bohrium\textsuperscript{\textregistered}(\url{https://bohrium.dp.tech}).

\bibliography{reference}
\bibliographystyle{unsrt}

\end{document}